\documentclass[journal,a4paper,onecolumn]{IEEEtran}
\IEEEoverridecommandlockouts

\usepackage{tikz}
\usepackage{epsfig}
\usepackage{epstopdf}
\usepackage{siunitx}
\DeclareSIUnit{\belmilliwatt}{Bm}
\DeclareSIUnit{\dBm}{\deci\belmilliwatt}
\usepackage{pbox}
\usepackage{makecell}
\usepackage{multirow}
\usepackage{amssymb}
\usepackage{amsmath}
\usepackage[normalem]{ulem}

\usepackage{blindtext}
\usepackage{etoolbox}
 
\makeatletter

\docsvlist{\@oddhead,\@evenhead,\ps@headings,\ps@IEEEtitlepagestyle,\ps@IEEEpeerreviewcoverpagestyle}
\makeatother

\usepackage{eso-pic}

\begin{document}

\title{Static and small-signal modeling of radiofrequency hexagonal boron nitride switches}
\author{Anibal Pacheco-Sanchez, Omar Jordán-García, Eloy Ramírez-García, David Jiménez
\thanks{This work has received funding from Instituto Politécnico Nacional SIP/ 20230362, from the European Union’s Horizon 2020 research and innovation programme under grant agreement No GrapheneCore3 881603 and from Ministerio de Ciencia, Innovación y Universidades under grant agreements RTI2018-097876-B-C21(MCIU/AEI/FEDER, UE), FJC2020-046213-I and PID2021-127840NB-I00 (MCIN/AEI/FEDER, UE). This  article  has been partially  funded  by  the  European  Union Regional  Development  Fund within the framework of the ERDF Operational Program of Catalonia 2014-2020 with the support of the Department de Recerca i Universitat, with a grant of 50\% of total cost eligible. GraphCAT project reference: 001-P-001702. (\textit{Corresponding authors: E. Ramírez-García and A. Pacheco-Sanchez}) \newline \indent A. Pacheco-Sanchez and D. Jiménez are with the Departament d'Enginyeria Electr\`{o}nica, Escola d'Enginyeria, Universitat Aut\`{o}noma de Barcelona, Bellaterra 08193, Spain. Email: AnibalUriel@Pacheco.uab.cat \newline \indent O. Jordán-García and E. Ramírez-García are with Instituto Politécnico Nacional, UPALM, Edif. Z-4 3er Piso. Email: ramirezg@ipn.mx}
}

\maketitle
\makeatletter
\def\ps@IEEEtitlepagestyle{
  \def\@oddfoot{\mycopyrightnotice}
  \def\@evenfoot{}
}
\def\mycopyrightnotice{
  {\footnotesize
  \begin{minipage}{\textwidth}
  \centering
© 2023 IEEE. Personal use of this material is permitted. Permission from IEEE must be obtained for all other uses, in any current or future media, including reprinting/republishing this material for advertising or promotional purposes, creating new collective works, for resale or redistribution to servers or lists, or reuse of any copyrighted component of this work in other works. DOI: DOI: 10.1109/JEDS.2023.3268349
  \end{minipage}
  }
}
\begin{abstract}
\boldmath
A first modeling approximation of the general performance of radiofrequency (RF) switches based on hexagonal boron nitride (hBN), a two-dimensional (2D) dielectric material is presented. The \textit{I-V} characteristics intrinsic and extrinsic impedance parameters, the return loss, insertion loss and isolation of RF 2D switches fabricated with hBN are described here by a equivalent circuit models. Straightforward analytical expressions are obtained. In contrast to conventional switches, the unique RF performance of the hBN switch, at ON-state, i.e., a direct improvement with frequency of the insertion loss, is accurately described by considering a capacitor in the intrinsic part of the model. The latter is suggested to be related to storaged charge during the resistive switching mechanism. The highest mean relative error obtained between modeling and measurements of the return loss is of \SI{7.6}{\%} with the approach presented here which overcomes the \SI{42.5}{\%} of difference obtained with a previous model with an incomplete intrinsic device description.
\end{abstract}
%
%\pacs{}% insert suggested PACS numbers in braces on next line
%
%\maketitle %\maketitle must follow title, authors, abstract and \pacs

\IEEEpeerreviewmaketitle
\section{Introduction}
\label{ch:intro}

Memristive non-volatile devices have gained attraction for low-power high-performance applications in both digital and analog scenarios \cite{LanSeb22}. Specifically for the latter kind of applications, two-dimensional (2D) radiofrequency (RF) switches have emerged as attractive competitors to conventional solid-state devices due to their low-power consumption, reduced footprint, THz cutoff frequencies and resistive switching features \cite{QiaTya16}, \cite{WaiAda21}. Molybdenium disulfide (MoS$_2$) and hexagonal boron nitride (hBN) are the 2D materials used so far to fabricate these novel RF switches \cite{GeWu18}-\cite{YanDah23} in a vertical metal-insulator-metal (MIM) configuration (cf. Fig \ref{fig:DC_EC}(a)). Specifically, 2D RF switches outperform RF switches based on memristive, micro-electromechanical system and phase change conventional (not 2D) technologies with similar output power ($\sim\SI{20}{\dBm}$) in terms of device area, switching time, cutoff frequency and thermal budget \cite{KimPal20}, \cite{KimDuc22}. hBN RF switches have been proven to work properly in high-data rate communication systems, e.g., for real-time video streaming \cite{KimPal20}.

Furthermore, a unique feature of hBN switches is the improvement of the insertion loss with frequency \cite{KimPal19,KimPal20} due to the dielectric nature of this 2D material \cite{LanVan18,MaPra18}. The latter suggests a non-negligible capacitance in the ON-state of 2D RF switches which is missing in the analytical expressions \cite{KimGe18}, \cite{KimPal20} and circuit simulator-based results \cite{KimPal19} available in the literature for these devices. In this work, the static and dynamic performance of 2D hBN RF switches is described accurately for the first time simultaneously. In section II,  the general modeling framework, i.e., theory and equivalent circuits, used to describe the \textit{I-V} curves as well as both the extrinsic and intrinsic insertion loss and isolation, i.e., $S_{21}$ at ON- and OFF-state, respectively. Section III presents the good agreement between the modeling approached used here and the experimental data reported in previous literature. Finally, general conclusions  for this work are included in the last section.

\section{Modeling framework}

\subsection{DC performance}

\begin{figure}[!hbt]
\centering
		\includegraphics[height=0.35\textwidth]{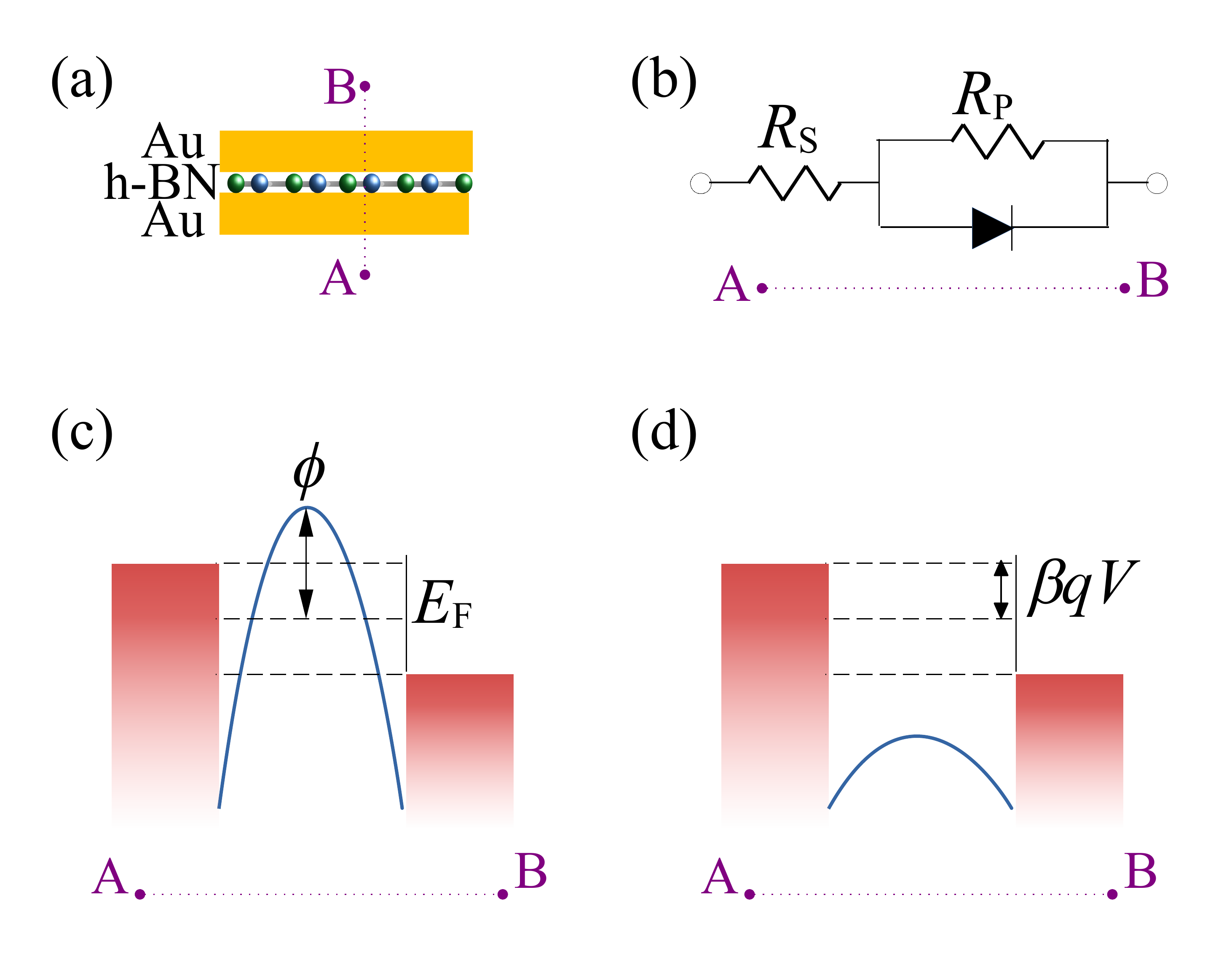}
\caption{hBN nonvolatile device. (a) Schematic cross-section of a hBN MIM-like structure. (b) DC equivalent circuit of the hBN nonvolatile device described by Eq. (\ref{eq:current00}). Energy diagrams of the device showing (c) a narrow and a (d) wide constriction, at HRS and LRS, respectively. $E_{\rm F}$ is the equilibrium Fermi energy; other variables are defined in the main text. The blue solid-line represents a potential barrier due to confinement effects considering the narrowest point of the filaments at the middle of the hBN layer. A-B dashed line is added to ease the discussion regarding the direction of the transport.}
\label{fig:DC_EC}
\end{figure}

Based on atomistic simulations and measurements, the transport during the different resistive switching states of hBN MIM-like devices (cf. Fig. \ref{fig:DC_EC}(a)) has been associated to the formation of conducting filaments enabled by boron vacancies \cite{KimPal20}, \cite{PanMir17}-\cite{MitMah22}. The electronic band diagram of the device is schematically represented by Figs. \ref{fig:DC_EC}(c) and (d). At the high-resistance state (HRS) the conducting filaments form a narrow path and hence a potential barrier due to the confinement effect on electrons passing through it arises (cf. Fig. \ref{fig:DC_EC}(c)). On the other hand, a wide constriction implies no barrier for the electrons as shown in Fig. \ref{fig:DC_EC}(d), i.e., a low-resistance state (LRS) is considered.

A practical description of the scenario described above is provided by the quantum point contact model based on the Landauer transport equation \cite{RuiMor16}. Hence, the current-voltage characteristics of hBN switches is described here by a Landauer-based compact model given by \cite{PanMir17}, \cite{RuiMor16}

\begin{equation}
 I = G_{\rm p}(V-IR_{\rm s}) + I_0\lbrace \exp \left[\alpha q (V - I R_{\rm{s}})\right] -1\rbrace,
 \label{eq:current00}
\end{equation}

\noindent where $V$ is the applied voltage, $G_{\rm p}=G_{\rm o}n$ with $G_{\rm o}$ as the quantum conductance ($\approx \SI{77.5}{\micro\siemens}$ for one transmission mode) and $n$ as the number of fully formed conducting filaments; $I_0 = [2q(N-n)/(\alpha h)]\exp(-\alpha \phi)$ with $q$ as the element charge, $N$ as the number of independent conducting filaments, $h$ as the Planck's constant and $\alpha$ and $\phi$ are parameters related to the width and height, respectively, of a confinement-related potential barrier \cite{PanMir17} (cf. Fig. \ref{fig:DC_EC}(c)); $R_{\rm s}$ is a series resistance embracing non-broken hBN layers, metal-contact layers and extrinsic phenomena. Eq. (\ref{eq:current00}) is a trascendental function for $I$ and it can be related to an equivalent circuit (EC) conformed by a resistance (first term) in parallel with a diode (second term) and both in series with $R_{\rm s}$ as seen in Fig. \ref{fig:DC_EC}(b).

The Thévenin transformation of the EC yields a solution for Eq. (\ref{eq:current00}) such as \cite{PanMir17}, \cite{RuiMor16}

\begin{equation}
 I_{\rm HRS} = \frac{VG_{\rm{p}}-I_0}{G_{\rm{A}}}+\frac{1}{\Lambda} W\Biggl\{ \frac{\Lambda I_0}{G_{\rm A}} \exp \left[\frac{\alpha q (V+R_{\rm s}I_0)}{G_{\rm A}} \right] \Biggl\},
\label{eq:current}
\end{equation}

\noindent where $G_{\rm{A}}=1+R_{\rm{s}}G_{\rm{p}}$, $\Lambda = \alpha q R_{\mathsf{s}}$ and $W\lbrace \Xi \rbrace$ is the Lambert function. Details on the derivation of the model have been provided elsewhere \cite{PanMir17}, \cite{RuiMor16}. Notice that Eq. (\ref{eq:current}) is only valid if the intrinsic potentials are lower than the barrier height, i.e., $q(V-IR_{\rm s})<\phi$ and if an asymmetric constriction is considered\footnote{This implies that the applied bias drops entirely at the source side of the constriction} \cite{PanMir17}. 

Eq. (\ref{eq:current}) describes the current at HRS where both effects, a resistance related to the the 2D material and the transport through a potential barrier, are present. However, for LRS, the latter effect has no impact on the current ($\phi\ll0$, see Fig. \ref{fig:DC_EC}(d)) due to the formation of a very wide energy constriction, i.e., $n\approx N$, and hence Eq. (\ref{eq:current00}) reduces to \cite{PanMir17}

\begin{equation}
I_{\rm LRS} = \frac{NG_{\rm o}}{1+NG_{\rm o}R_{\rm s}} V.
\label{eq:current_lrs}
\end{equation}

This modeling approach has been used before to describe the resistive states of hBN nonvolatile devices \cite{PanMir17}, \cite{MirSun17}-\cite{VilHui19}, however, this is the first time it is applied to model the DC performance of hBN devices intended for state-of-the-art RF applications.

\subsection{Dynamic response}

\begin{figure}[!hbt]
\centering
		\includegraphics[height=0.45\textwidth]{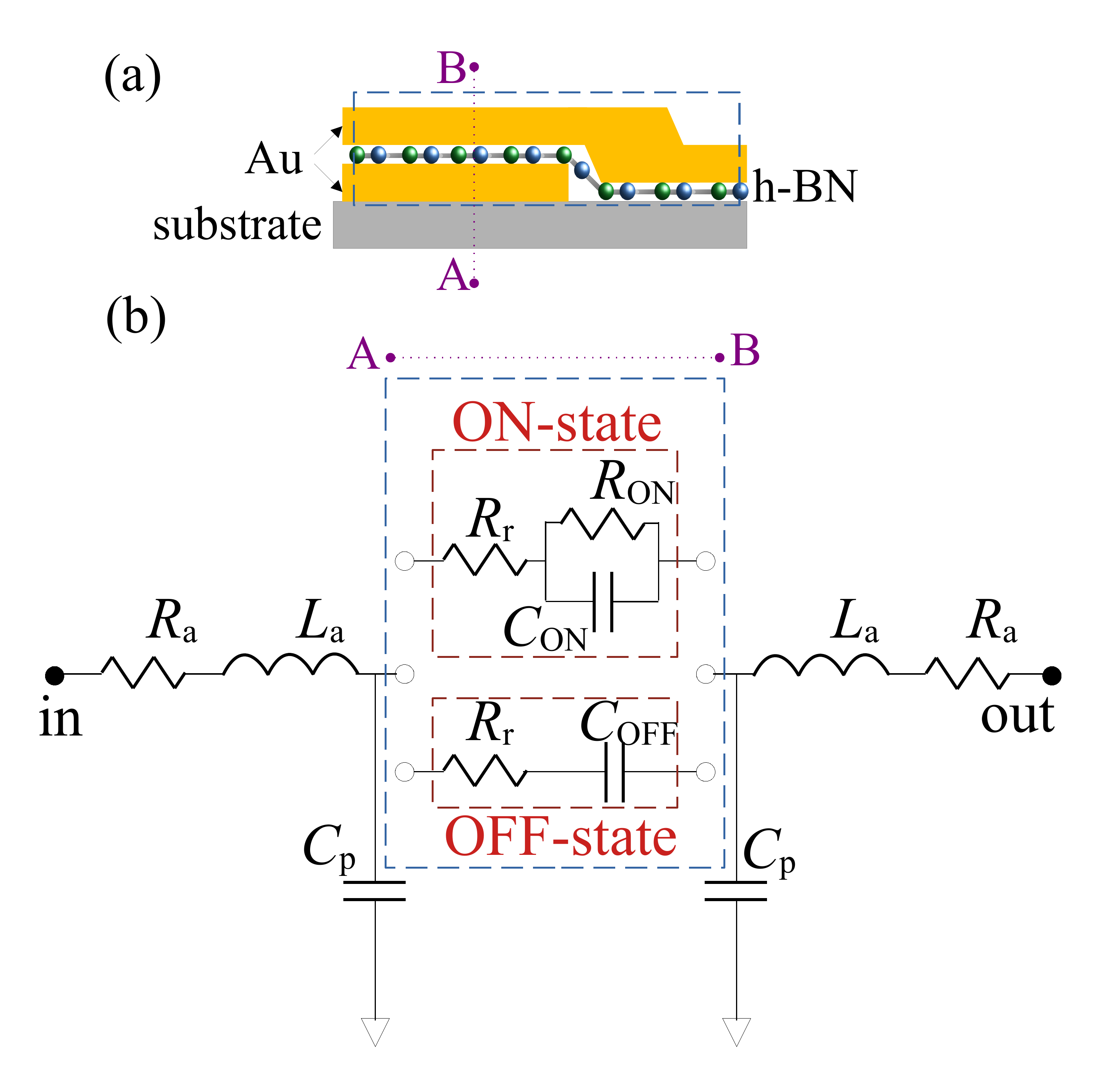}
\caption{(a) Schematic cross section of an h-BN RF switch; intrinsic device is inside the dashed box whereas extrinsic part is not shown. (b) Equivalent circuit of an RF switch. Intrinsic (extrinsic) elements are inside (outside) the dashed box. A switch at ON-(OFF-)state is represented by the top (bottom) intrinsic impedance block. The intrinsic elements correspond to the intrinsic device according to the AB pointed line.}
\label{fig:EC}
\end{figure}

The cross-section of an hexagonal borond nitride RF switches, similar as the ones  demonstrated in the literature \cite{KimPal20}, \cite{KimDuc22}, \cite{YanDah23}, is shown in Fig. \ref{fig:EC}(a). As a first approximation for the description of the high-frequency performance of 2D RF switches, the small-signal EC shown in Fig. \ref{fig:EC}(b) has been used. The intrinsic part of the EC changes according to the switch state and it can be derived from the circuit shown in Fig. \ref{fig:DC_EC}(b) by considering the small-signal EC of the diode \cite{CowSor66}. For the ON-state, a parallel array of a resistance $R_{\rm ON}$ and a capacitor $C_{\rm ON}$ have been used to represent the resistance and a non-negligible charge, respectively, within the activated MIM structure. The latter arrangement is obtained by considering the small-signal EC of the diode (cf. Fig. \ref{fig:DC_EC}(b)) and hence $R_{\rm ON}$ is approximately equal to the voltage-dependent resistance of the diode\footnote{This approximation is guaranteed by $R_{\rm P}$ ($\approx \SI{12.9}{\kilo\ohm}$) being much higher than the voltage-dependent diode resistance (with a lower value) to which it is in parallel with.}. A capacitor $C_{\rm OFF}$ represents the electrostatic coupling between device materials in the OFF-state. For both states in the intrinsic part, a residual series resistance $R_{\rm r}$ has been included here in contrast to previous works in the literature \cite{KimPal19,KimPal20}, \cite{YanMak22}, \cite{JorRam22} where such effect has been neglected. For the ON-state, i.e., at LRS, $R_{\rm r} = R_{\rm S}$, whereas at the OFF-state, i.e., at HRS, $R_{\rm r}$ is a fitting parameter. Two aspects should be noticed for  $R_{\rm r}$ at OFF-state: \textit{(i)} it is required to prevent infinite isolation at low frequency, and \textit{(ii)} the value differs from the one used for $R_{\rm S}$ in Eq. (\ref{eq:current}) due to a possible independent measurement of DC and RF characteristics, i.e., different cycles might be used to obtain the corresponding characteristics. The extrinsic network is comprised by a parasistic capacitor $C_{\rm p}$, an access resistance $R_{\rm a}$ and an inductance $L_{\rm a}$ associated to each port.

The $S_{21}$-parameter for a two-port reciprocal and symmetrical network, such as the one describing the RF nanoswitch (c.f. Fig. \ref{fig:EC}), in terms of the $Z$-parameters is given by \cite{Poz05}

\begin{equation}
S_{\rm 21,ext} = \frac{2Z_{\rm 21,ext}Z_0}{\left(Z_{\rm 11,ext}+Z_0\right)^2-Z_{\rm 21,ext}^2},
\label{eq:s21_ext}
\end{equation}

\noindent where $Z_0=\SI{50}{\ohm}$ is the characteristic impedance, whereas the impedance terms for the device with extrinsic contributions are 

\begin{equation}
%Z_{\rm 11,ext} = \frac{V_{\rm in}}{I_{\rm in}}\bigg\vert_{I_{\rm out}=0} = Z_{\rm a} + Z_{\rm \alpha} \left(Z_{\rm int} + Z_{\rm c}\right),
Z_{\rm 11,ext} = Z_{\rm a} + Z_{\rm \alpha} \left(Z_{\rm int} + Z_{\rm c}\right),  
\label{eq:z11_ext}
\end{equation}

\noindent and

\begin{equation}
%Z_{\rm 21,ext} = \frac{V_{\rm out}}{I_{\rm in}}\bigg\vert_{I_{\rm out}=0} = Z_{\alpha}  Z_{\rm c} ,
Z_{\rm 21,ext} = Z_{\alpha}  Z_{\rm c} ,
\label{eq:z21_ext}
\end{equation}

\noindent with 

\begin{equation}
Z_{\alpha} = \frac{Z_{\rm c}}{Z_{\rm int} + 2Z_{\rm c}},
\end{equation}

\noindent the access impedance $Z_{\rm a} = R_{\rm a} + j\omega L_{\rm a}$, the extrinsic impedance due to pads electrostatic coupling $Z_{\rm c}=1/(j\omega C_{\rm p})$ and the intrinsic impedance $Z_{\rm int}$ as

\begin{equation}
Z_{\rm int} = \begin{cases} R_{\rm r} + \frac{R_{\rm ON}}{1+j\omega R_{\rm ON}C_{\rm ON}} &\mbox{for ON state,} \\
R_{\rm r} + \frac{1}{j\omega C_{\rm OFF}} & \mbox{for OFF state,} \end{cases}
\label{eq:zint}
\end{equation}

\noindent where $\omega=\SI{2}{}\pi f$ is the angular frequency with $f$ as the operation frequency. By using Eqs. (\ref{eq:z11_ext})-(\ref{eq:zint}), and the corresponding values of the elements, in Eq. (\ref{eq:s21_ext}), the extrinsic insertion loss (ON-state) and isolation (OFF-state) are obtained. Notice that by doing the corresponding parameter transformation \cite{Poz05}, Eq. (\ref{eq:s21_ext}) is equivalent to the one given for conventional lateral RF switches considering \textit{ABCD} parameters \cite{Liu10}\footnote{The expressions reported for $S_{\rm 21,ext}$, obtained with \textit{ABCD} parameters, in \cite{KimPal19} and \cite{KimPal20} have both a typo in the second term of the denominator}.

On the other hand, the intrinsic part of the 2D RF switch can be considered as a series impedance block and hence, the intrinsic insertion loss and isolation can be obtained by \cite{Cas10}

\begin{equation}
S_{\rm 21,int} = \frac{2}{2+\frac{Z_{\rm int}}{Z_0}},
\label{eq:s21_int}
\end{equation}

\noindent depending on whether $Z_{\rm int}$ is defined for the ON- or OFF-state, respectively. Notice that the approach presented here considers the contribution of $C_{\rm ON}$ in the expressions of both extrinsic and intrinsic insertion loss in contrast to previous studies \cite{KimGe18}-\cite{KimPal20} where this effect has been neglected. 

The extrinsic and intrinsic return loss $RL=-20\log S_{11}$ \cite{Poz05}, considering the EC in Fig. \ref{fig:EC} and Eqs. (\ref{eq:z11_ext})-(\ref{eq:zint}), can be calculated from

\begin{equation}
S_{\rm 11,ext} = \frac{Z_{11,ext}^2-Z_0^2-Z_{\rm 21,ext}^2}{(Z_{\rm 11,ext}+Z_0)^2-Z_{\rm 21,ext}^2},
\label{eq:s11ext}
\end{equation}

\noindent and 

\begin{equation}
S_{\rm 11,int} = \frac{1}{1+\frac{2Z_0}{Z_{\rm int}}},
\label{eq:s11int}
\end{equation}

\noindent respectively, at ON-state.

\section{Results and discussion}

\begin{figure}[!htb]
\centering
\includegraphics[width=0.45\textwidth]{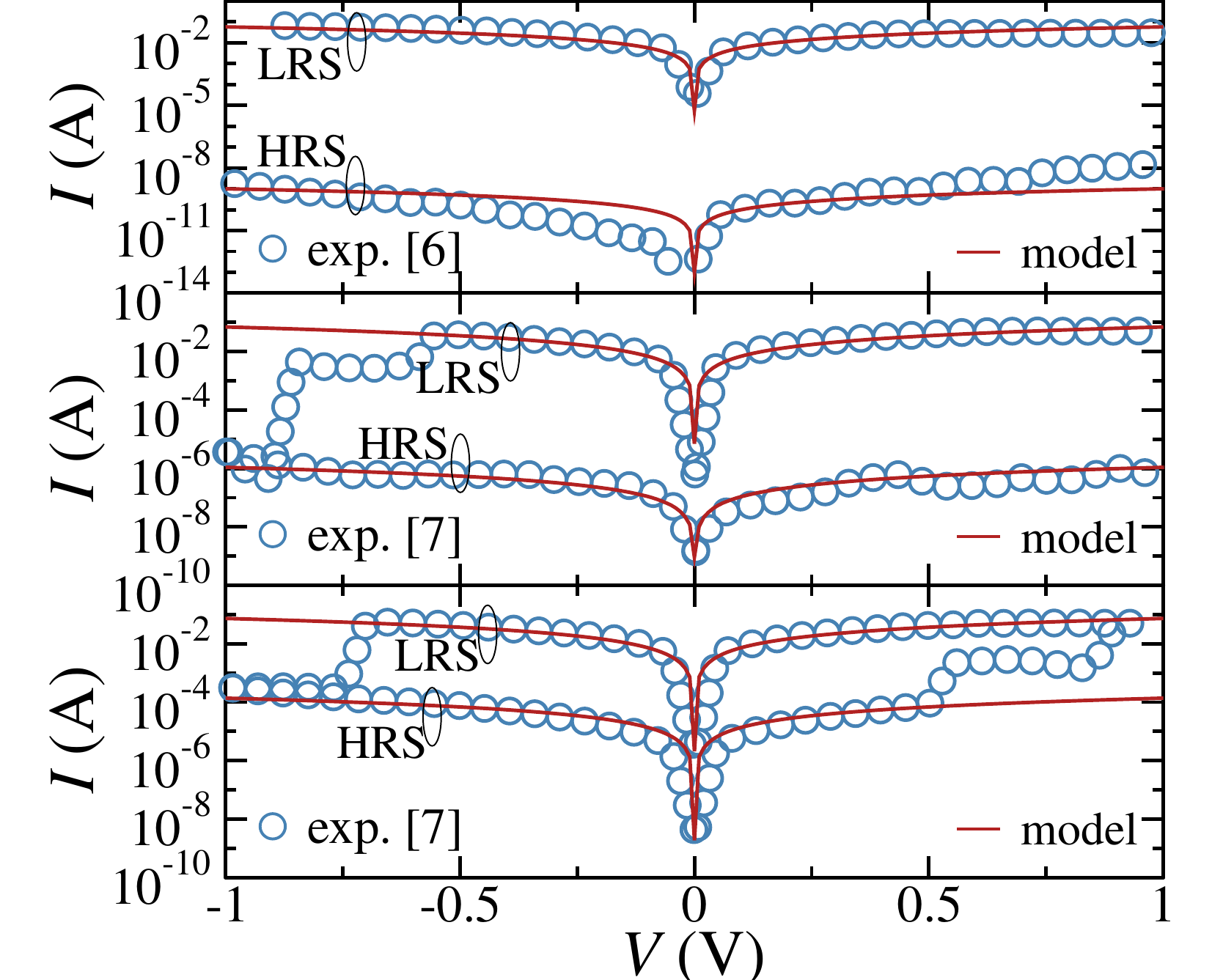}
\caption{\textit{IV} curves of hBN RF switches with different device areas: (top) $\SI{0.50}{}\times\SI{0.15}{\micro\meter^2}$ \cite{KimPal19}, (middle) $\SI{0.20}{}\times\SI{0.15}{\micro\meter^2}$ \cite{KimPal20} and (bottom) $\SI{0.50}{}\times\SI{0.50}{\micro\meter^2}$ \cite{KimPal20}. LRS model parameters: (top) $N=\SI{740}{}$, $R_{\rm s}=\SI{9.5e-9}{\ohm}$; (middle) $N=\SI{900}{}$, $R_{\rm s}=\SI{9.5e-5}{\ohm}$; (bottom) $N=\SI{950}{}$, $R_{\rm s}=\SI{5e-6}{\ohm}$. HRS model parameters: (top) $N=\SI{740}{}$, $n=\SI{3}{}$, $R_{\rm s}=\SI{1e9}{\ohm}$, $\alpha=\SI{4.6}{\electronvolt^{-1}}$, $\phi=\SI{6.5}{\electronvolt}$; (middle) $N=\SI{900}{}$, $n=\SI{1}{}$, $R_{\rm s}=\SI{9e5}{\ohm}$, $\alpha=\SI{3.6}{\electronvolt^{-1}}$, $\phi=\SI{5.5}{\electronvolt}$; (bottom) $N=\SI{950}{}$, $n=\SI{17}{}$, $R_{\rm s}=\SI{6.5e3}{\ohm}$, $\alpha=\SI{3.9}{\electronvolt^{-1}}$, $\phi=\SI{5.4}{\electronvolt}$.}
\label{fig:iv}
\end{figure}

The modeling approach presented above has been applied to describe the experimental \textit{I-V} performance, RF insertion loss and RF isolation of fabricated hBN RF switches \cite{KimPal19,KimPal20} with different device footprints. Details on the fabrication processes and measurements have been presented elsewhere \cite{KimPal19,KimPal20}. \textit{I-V} characteristics of hBN RF switches have been successfully described by the model at both states, i.e., HRS and LRS by Eqs. (\ref{eq:current}) and (\ref{eq:current_lrs}), respectively, as shown in Fig. \ref{fig:iv}. Notice that the intrinsic voltage is lower than the barrier height in all devices, i.e., Eq. (\ref{eq:current}) is valid. The large difference in the value of $R_{\rm s}$ at ON-state (LRS) and OFF-state (HRS) is due to the potential step affecting the transport differently at each case.
%Notice that the asymmetry of the experimental data (around \SI{0}{\volt}) of \cite{KimPal19} at HRS is not useful for practical applications and hence, it is out of the scope of the model used here. The latter reveals the case of practical symmetric \textit{IV} data.

Regarding the small-signal response, Table \ref{tab:values} collects the intrinsic parameter values used here for describing the experimental data of 2D RF switches. Values of $R_{\rm ON}$ and $C_{\rm OFF}$ obtained here are in good agreement with the reported ones in \cite{KimPal19} and for the smallest device in \cite{KimPal20}.  For the largest device in \cite{KimPal20}, $R_{\rm ON}$ and $C_{\rm OFF}$ have been obtained by considering a correct description of the RF power handling feature of the switch. $C_{\rm ON}$ values have not been reported previously for ECs of hBN switches in the literature and are obtained here by considering a correct description of $S_{\rm 21,int}$ in the ON-state for each switch. The values of $C_{\rm ON}$ obtained for the hBN devices are similar to the ones reported elsewhere for fabricated hBN-based MIMs, i.e., few \SI{}{\pico\farad} \cite{AhmHeo18,HeLi21}. Furthermore, this capacitive effect at ON-state has not been observed in other 2D monolayer RF switches \cite{GeWu18}, \cite{KimGe18} and hence it is suggested here to be associated to the charge storaged within the hBN layer during the filament-based resistive switching \cite{WuGe19} rather than to the thickness of the monolayer as suggested elsewhere \cite{KimPal19}. A further discussion of the intrinsic mechanism of the resisitive switching mechanism in hBN is out of the scope of this work, however, interested readers are directed to \cite{WuGe19}.

\begin{table} [!htb] 
\begin{center}
\caption{Small-signal EC intrinsic parameter values of hBN RF switches.}
\begin{tabular}{c|c||c|c|c|c}
[ref.] & \makecell{device area \\ ($\SI{}{\micro\meter^2}$)} & \makecell{$R_{\rm r,ON/OFF}$\\(\SI{}{\ohm})} & \makecell{$R_{\rm ON}$\\(\SI{}{\ohm})} & \makecell{$C_{\rm ON}$\\($\SI{}{\pico\farad}$)} & \makecell{$C_{\rm OFF}$\\(\SI{}{\femto\farad)}}  \\ \hline \hline

%\makecell{$\SI{0.50}{}\times\SI{0.50}{\micro\meter^2}$\\MoS$_2$ \cite{KimGe18}} & \SI{4.2}{} & -- & \SI{5.4}{} & ? & ? & ? \\ 

 \cite{KimPal19} & $\SI{0.50}{}\times\SI{0.15}{}$ & $\SI{9.5e-5}{}/\SI{50}{}$ & \SI{1.61}{} & \SI{1.00}{} & \SI{2.00}{} \\

\cite{KimPal20} & $\SI{0.20}{}\times\SI{0.15}{}$ & $\SI{9.5e-5}{}/\SI{900}{}$ & \SI{2.81}{} & \SI{1.21}{} & \SI{0.44}{} \\

 \cite{KimPal20} & $\SI{0.50}{}\times\SI{0.50}{}$ & $\SI{5e-6}{}/\SI{1800}{}$ & \SI{1.65}{} & \SI{1.21}{} & \SI{1.51}{} \\
 
\end{tabular} \label{tab:values}
\end{center}
\end{table}

\begin{figure}[!htb]
\centering
\includegraphics[height=0.2025\textwidth]{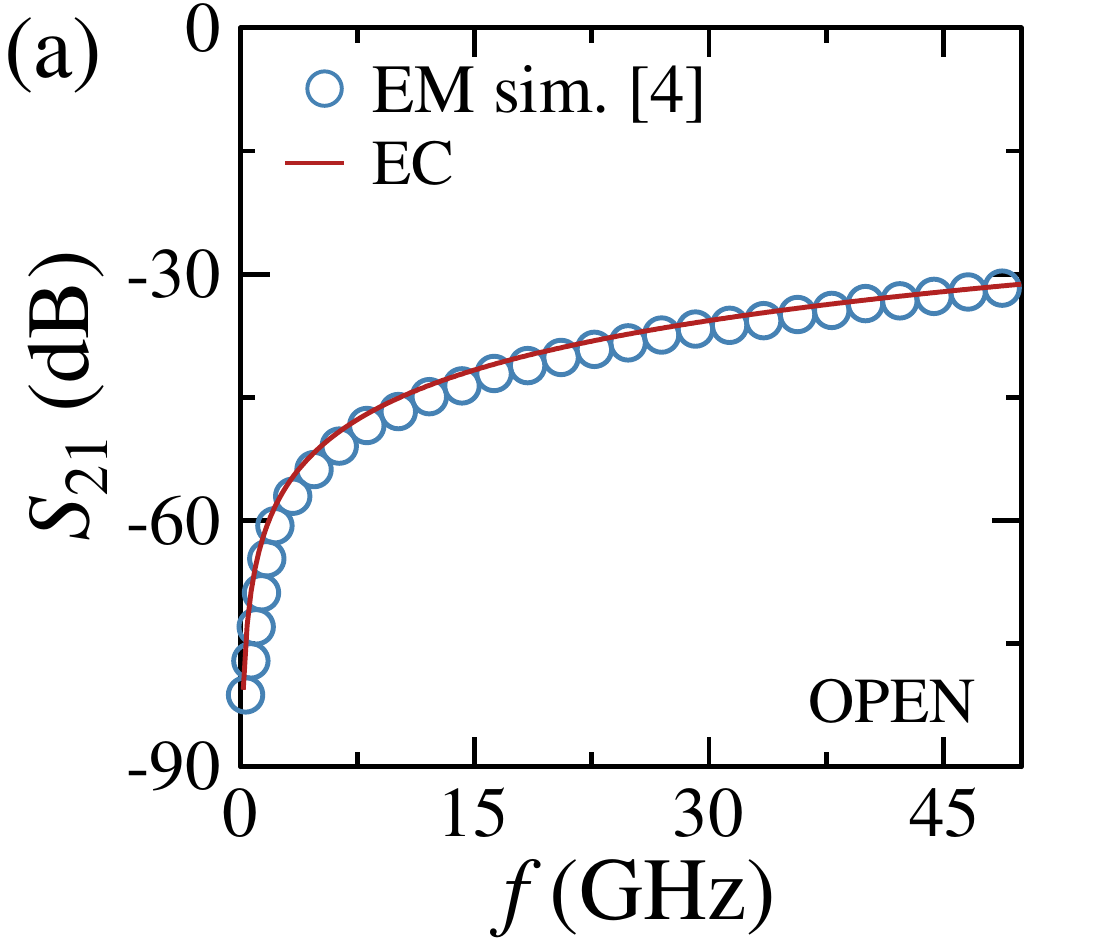}
\includegraphics[height=0.2025\textwidth]{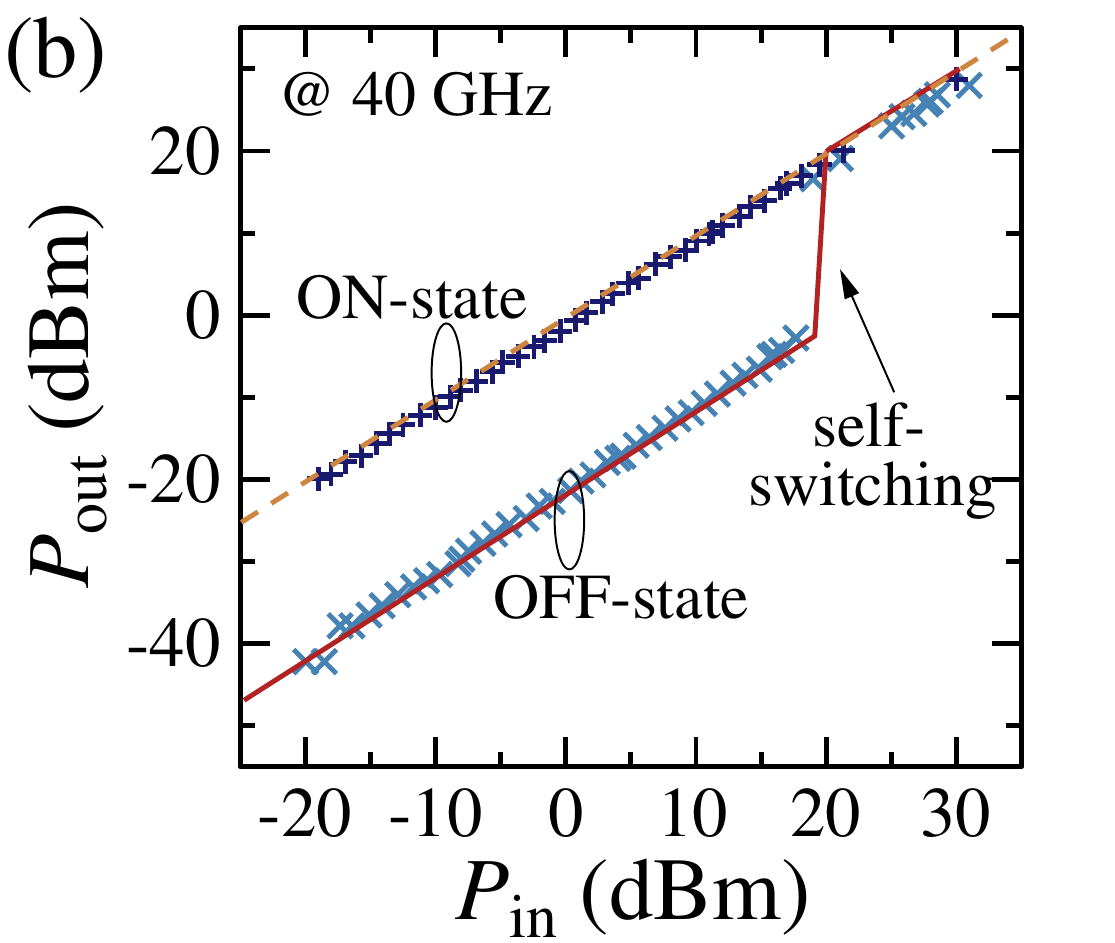} \\
\caption{(a) $S_{21}$ of an OPEN structure obtained with electromagnetic simulations presented in \cite{KimPal19} (symbols) and with the extrinsic network of the equivalent circuit presented here in Fig. \ref{fig:EC} (line). (b) Experimental (markers) and simulation results with the EC (lines) of the signal power handling of the $\SI{0.5}{}\times\SI{0.5}{\micro\meter^2}$ hBN switch at \SI{40}{\giga\hertz} presented in \cite{KimPal20}.}
\label{fig:calibration}
\end{figure}

The EC extrinsic network for all devices has been calibrated to the electromagnetic simulations of the OPEN-structure presented in \cite{KimPal19}. The obtained values of $C_{\rm p}$ and $R_{\rm a}$ (cf. Fig. \ref{fig:calibration}(a)) are equal to \SI{7.50}{\femto\farad} and \SI{1.25}{\ohm}, respectively. An $L_{\rm a}$ of \SI{0.25}{\pico\henry} corresponds to an average value of a minimum and a maximum limit of a gold access pad, such as the one used in \cite{KimPal19,KimPal20}, considering a coplanar waveguide and an isolated microstrip, respectively \cite{AnhSwi91}. 

The EC describes the experimental RF power handling in both ON- and OFF-state of the device in \cite{KimPal20} at \SI{40}{\giga\hertz}, as well as its self-switching, as shown in Fig. \ref{fig:calibration}(b). The self-switching from OFF- to ON-state is achieved with the EC by considering a transition from $C_{\rm OFF}$ to $C_{\rm ON}$ (see values in Table \ref{tab:values}) which, along with the accurate description of ON-state results, confirms the value proposed here for the latter element.

\begin{figure}[!htb]
\centering
\includegraphics[height=0.2025\textwidth]{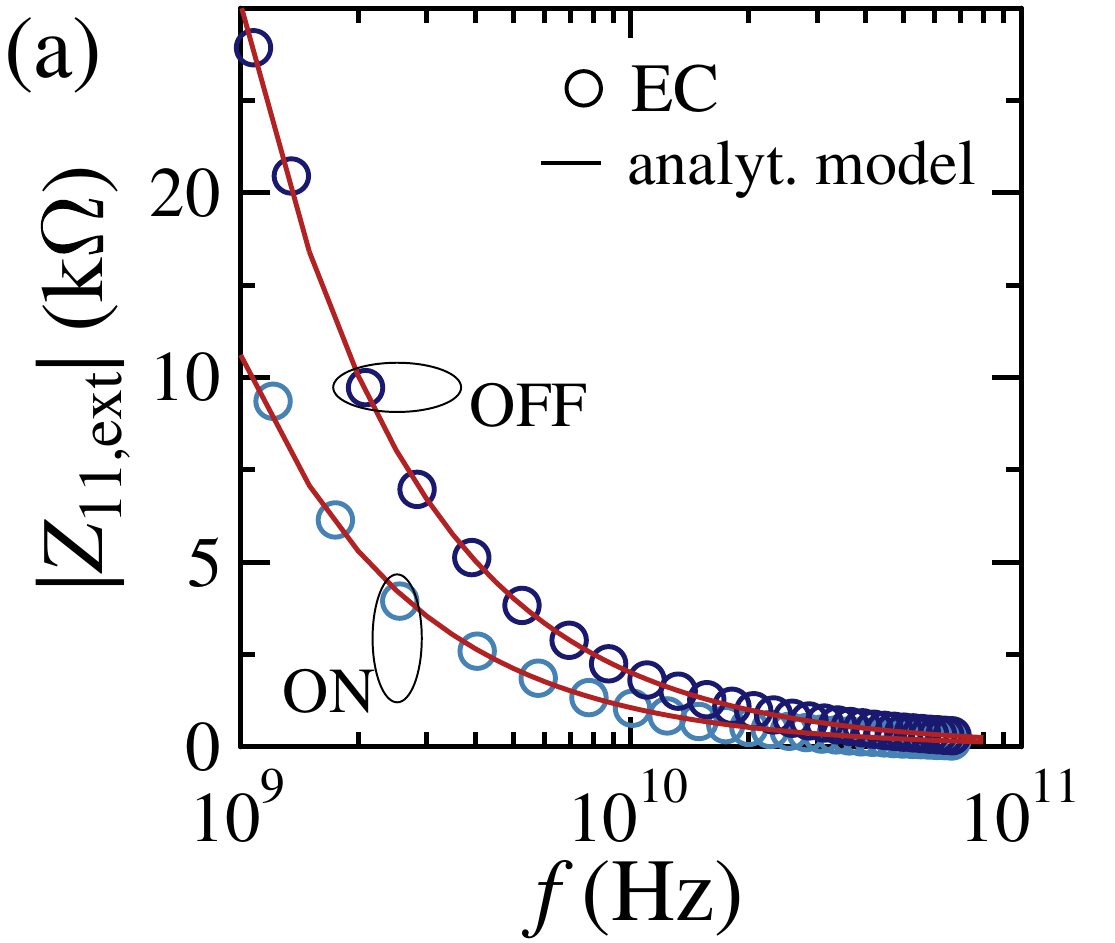}
\includegraphics[height=0.2025\textwidth]{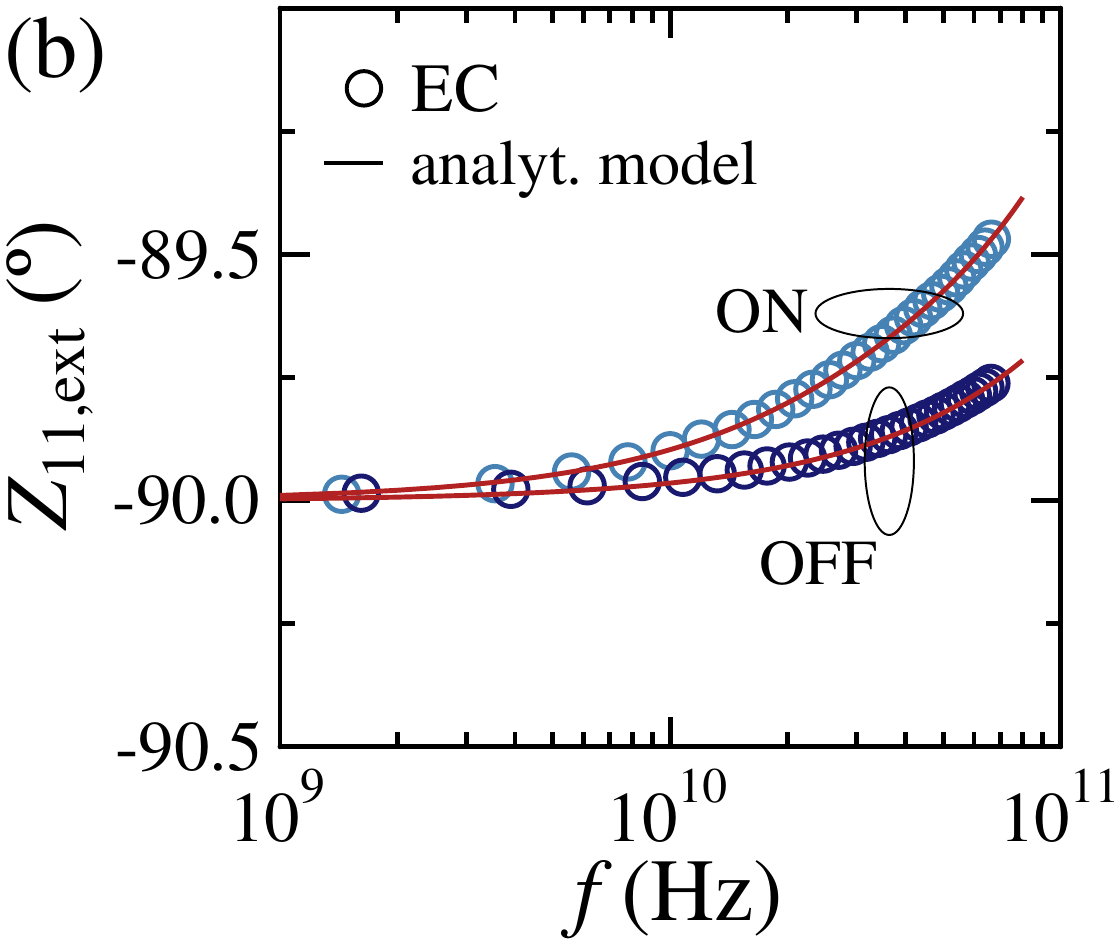} \\ 
\includegraphics[height=0.2025\textwidth]{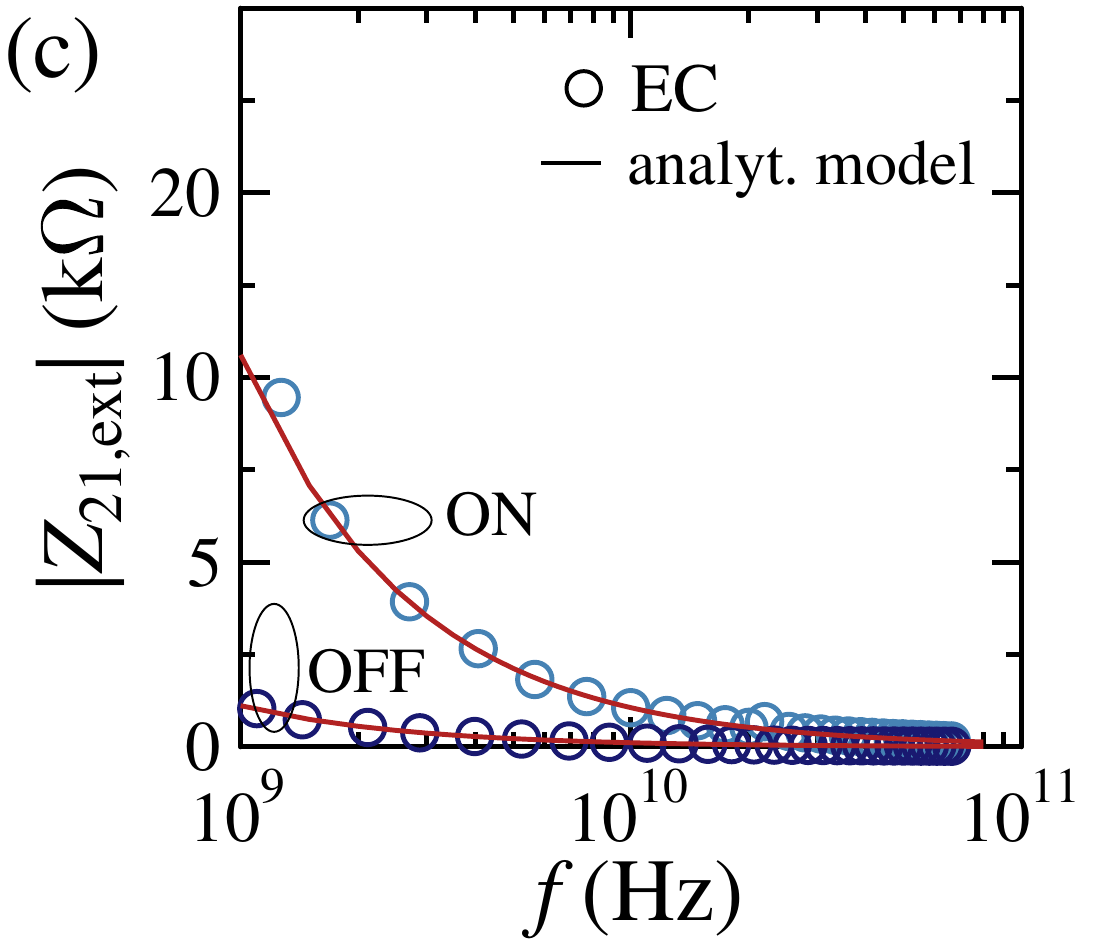}
\includegraphics[height=0.2025\textwidth]{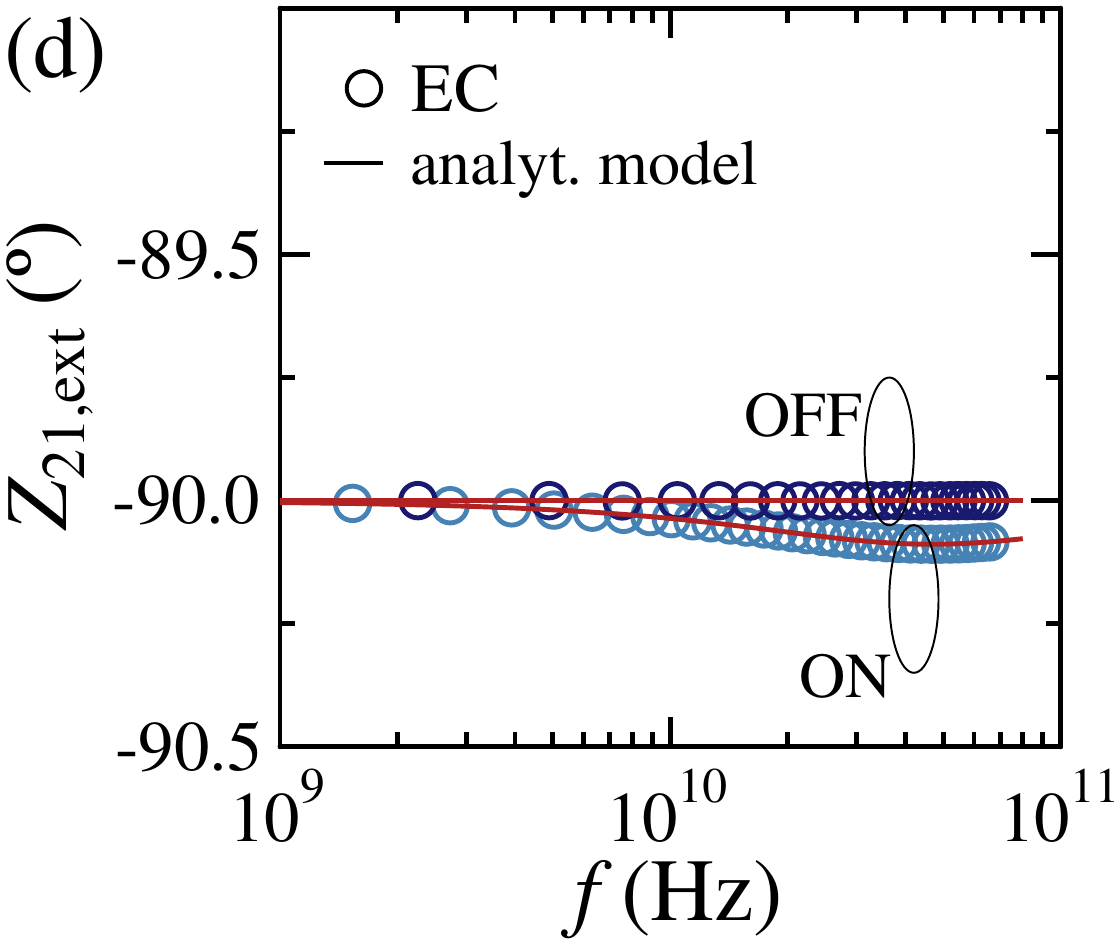}
\caption{Extrinsic impedance parameters over frequency considering the $\SI{0.20}{}\times\SI{0.15}{\micro\meter^2}$ device in \cite{KimPal20}. (a),(c)  Magnitude and (b),(d) phase of $Z_{11}$ and $Z_{21}$. Simulation results with the EC (markers) and modeling results (lines) with Eqs. (\ref{eq:z11_ext}) and (\ref{eq:z21_ext}).}
\label{fig:z_param}
\end{figure}

In order to validate the expressions considering extrinsic elements, a comparison with results of the experimentally-calibrated EC has been performed for the devices under study. Eqs. (\ref{eq:z11_ext}) and (\ref{eq:z21_ext}) reproduce properly the TCAD results of the impedance parameters (magnitude and phase) obtained here (with both extrinsic and intrinsic elements), regardless the state of the switch, for the smallest hBN device presented in \cite{KimPal20} as shown in Fig. \ref{fig:z_param}.  

\begin{figure}[!htb]
\centering
\includegraphics[width=0.45\textwidth]{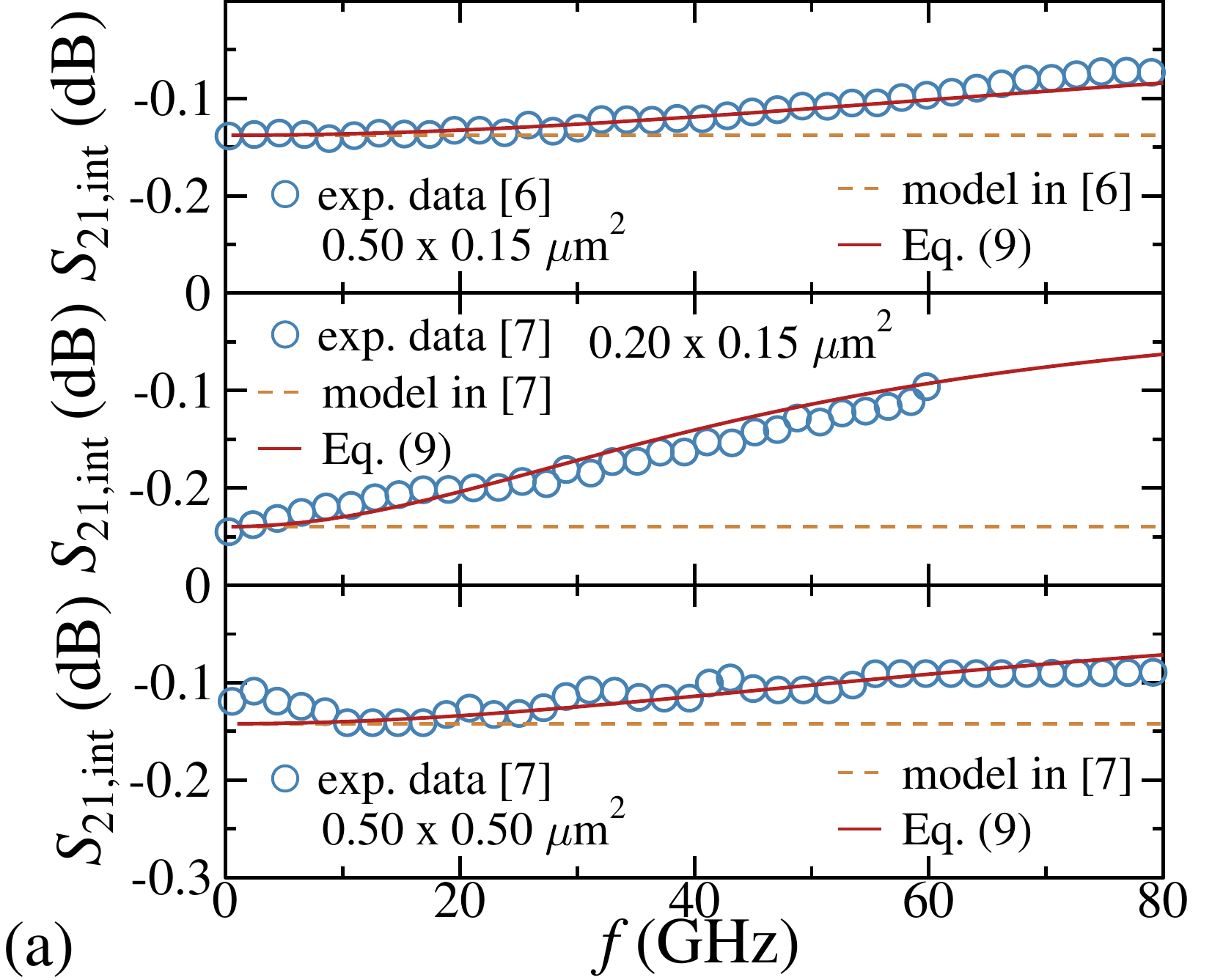} \\  \vspace{0.2cm}
\includegraphics[width=0.45\textwidth]{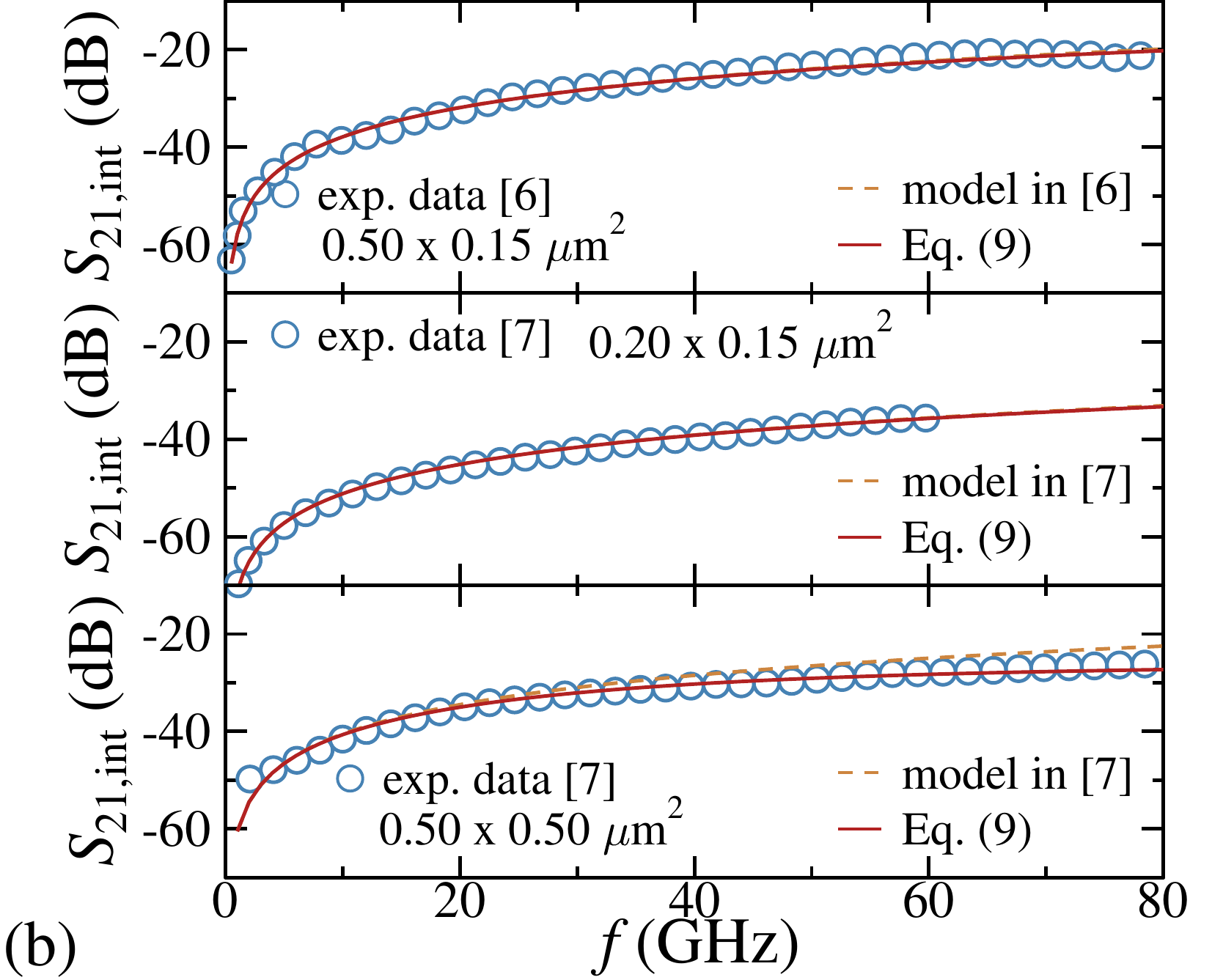}
\caption{Intrinsic (a) insertion loss and (b) isolation of 2D hBN RF switches with different area: $\SI{0.5}{}\times\SI{0.15}{\micro\meter^2}$ \cite{KimPal19} and $\SI{0.20}{}\times\SI{0.15}{\micro\meter^2}$ \cite{KimPal20}. Markers are experimental data. Dashed and solid lines are modeling results reported in the corresponding reference and with the approach presented here, respectively.}
\label{fig:results}
\end{figure}

The comparison between de-embedded (intrinsic) experimental $S_{\rm 21,int}$ and the different models, namely the one presented here (cf. Eq. (\ref{eq:s21_int})) and the one in the corresponding original reference \cite{KimPal19}, \cite{KimPal20}, is shown in Fig. \ref{fig:results}\footnote{Extrinsic elements are not considered in the experimental data nor in the modeling results.}. The modeling approach presented here differs from the previous reported model in \cite{KimPal19}, \cite{KimPal20} since the former one considers $C_{\rm ON}$ (for the ON-state) and $R_{\rm r}$ (for both states) in the EC in contrast to the latter one where both of these elements are neglected. 

For the device in \cite{KimPal19} and the smallest one in \cite{KimPal20}, the impact of $R_{\rm r}$ is minimum in the description of the insertion loss (cf. Fig. \ref{fig:results}(b)). However, for the largest device in \cite{KimPal20} the model used here including $R_{\rm r}$ fits better the experimental data in contrast to the lumped EC not considering this element (cf. Fig. \ref{fig:error}(b) in Appendix). On the other hand, for the ON-state, the increase of $S_{21,int}$ with frequency is not captured by the pure resistive intrinsic part of the models reported in \cite{KimPal19} and \cite{KimPal20}. Eq. (\ref{eq:s21_int}) obtained here is able to reproduce such unique attractive feature of RF hBN switches by considering a capacitance different to zero in the intrinsic part of the device (cf. Eq. (\ref{eq:zint})), in contrast to the conventional approaches developed for other switches not presenting such effect. The relative error $\epsilon_{\rm r}$ of each modeling approach with respect to the experimental data is shown in Fig. \ref{fig:error} in the Appendix where the higher accuracy of the model presented here with respect to others reported in the literature is observed, e.g.,the average (over the analyzed frequency range) of $\vert\epsilon_{\rm r}\vert$ (in percentage) for $S_{\rm 21,int}$ at ON-state is of \SIlist{6.9;7.6;1.8}{\%} with the model proposed here and \SIlist{27.6;42.5;9.9}{\%} with other approaches for the devices in \cite{KimPal19} and \cite{KimPal20}, respectively. 

\begin{figure}[!htb]
\centering
\includegraphics[height=0.225\textwidth]{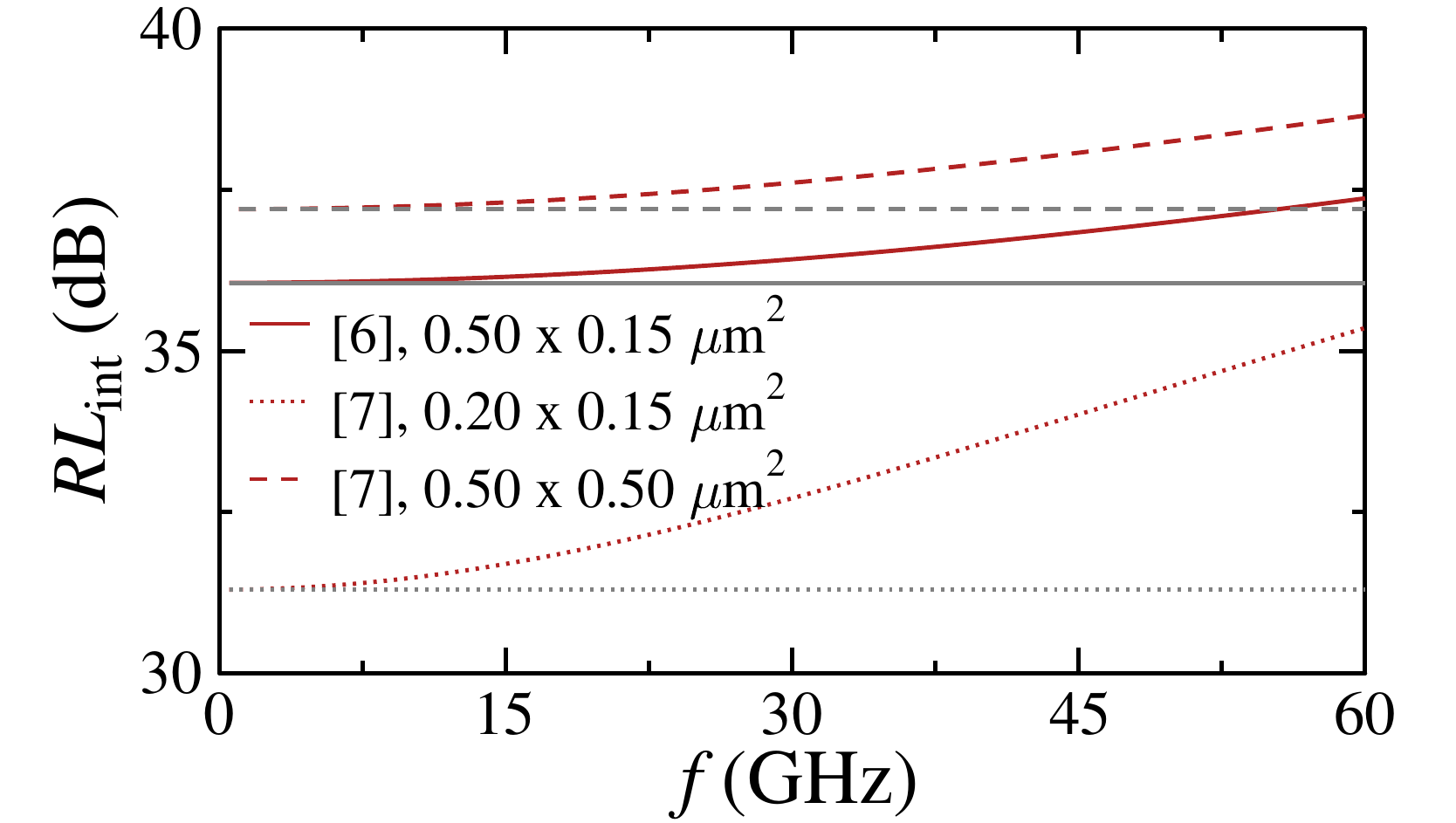}
\caption{Intrinsic return loss of 2D hBN RF switches obtained from Eq. (\ref{eq:s11int}). Red (gray) lines are results with (without) $C_{\rm ON}$.}
\label{fig:RL}
\end{figure}

The return loss, an important factor for RF switches \cite{Liu10}, not reported experimentally for any of the devices under study \cite{KimPal19}, \cite{KimPal20}, has been calculated here by using Eq. (\ref{eq:s11int}). Fig. \ref{fig:RL} shows that the $RL$ improves with the frequency due to the effect of the capacitance at ON-state whereas the device with largest area in \cite{KimPal20} has the highest $RL$ of the three hBN switches.

\section{Conclusion}

The static characteristics of hBN-based MIM-like structures, intended for RF applications, have been described for the first time here at both resistive states by using a physics-based equivalent circuit and the corresponding equations. Furthermore, the dynamic performance of 2D RF switches based on hBN has been precisely modeled here by considering an intrinsic charge contribution -via $C_{\rm ON}$- related to the storaged charge within this 2D material during the filament-based resistive switching mechanism. An analysis of the equivalent circuit considering such contribution yields straightforward expressions used to describe a unique increase with frequency of the $S_{21}$ parameter of RF hBN switches. Both the EC and the derived equations for impedance and scattering parameters are able to describe experimental data of three different devices. The relative error of the return loss and the insertion loss shows the higher accuracy of the approach discussed here with respect to other previously used models. The expressions obtained here for the dynamic performance of hBN switches aim to boost the modeling and analysis of these devices as well as their use in circuit design by considering a physical phenomenon represented by $C_{\rm ON}$, neglected in previous works, in their equivalent circuit.

\section*{Appendix}

\begin{figure}[!htb]
\centering
\includegraphics[width=0.45\textwidth]{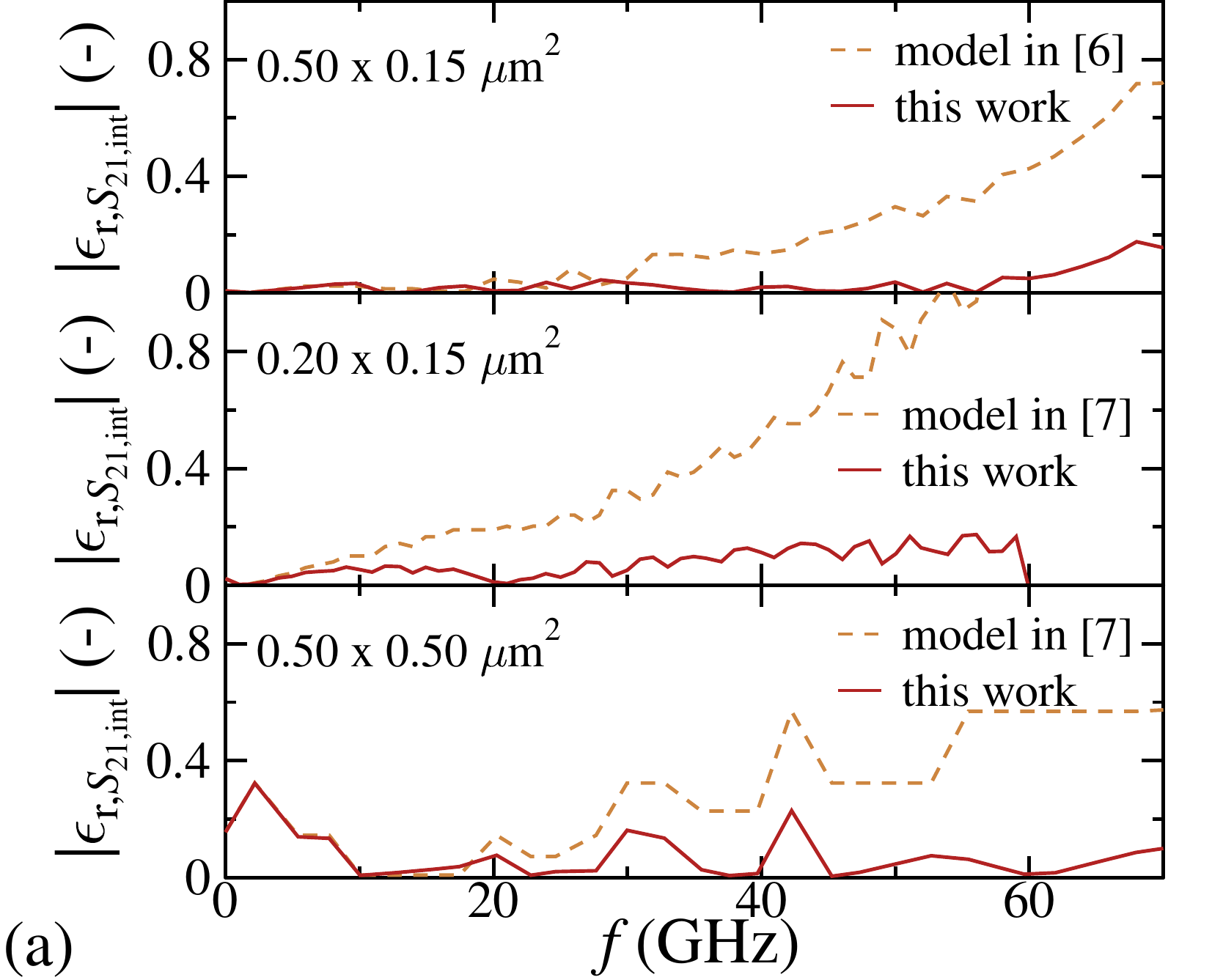} \\ \vspace{0.2cm}
\includegraphics[width=0.45\textwidth]{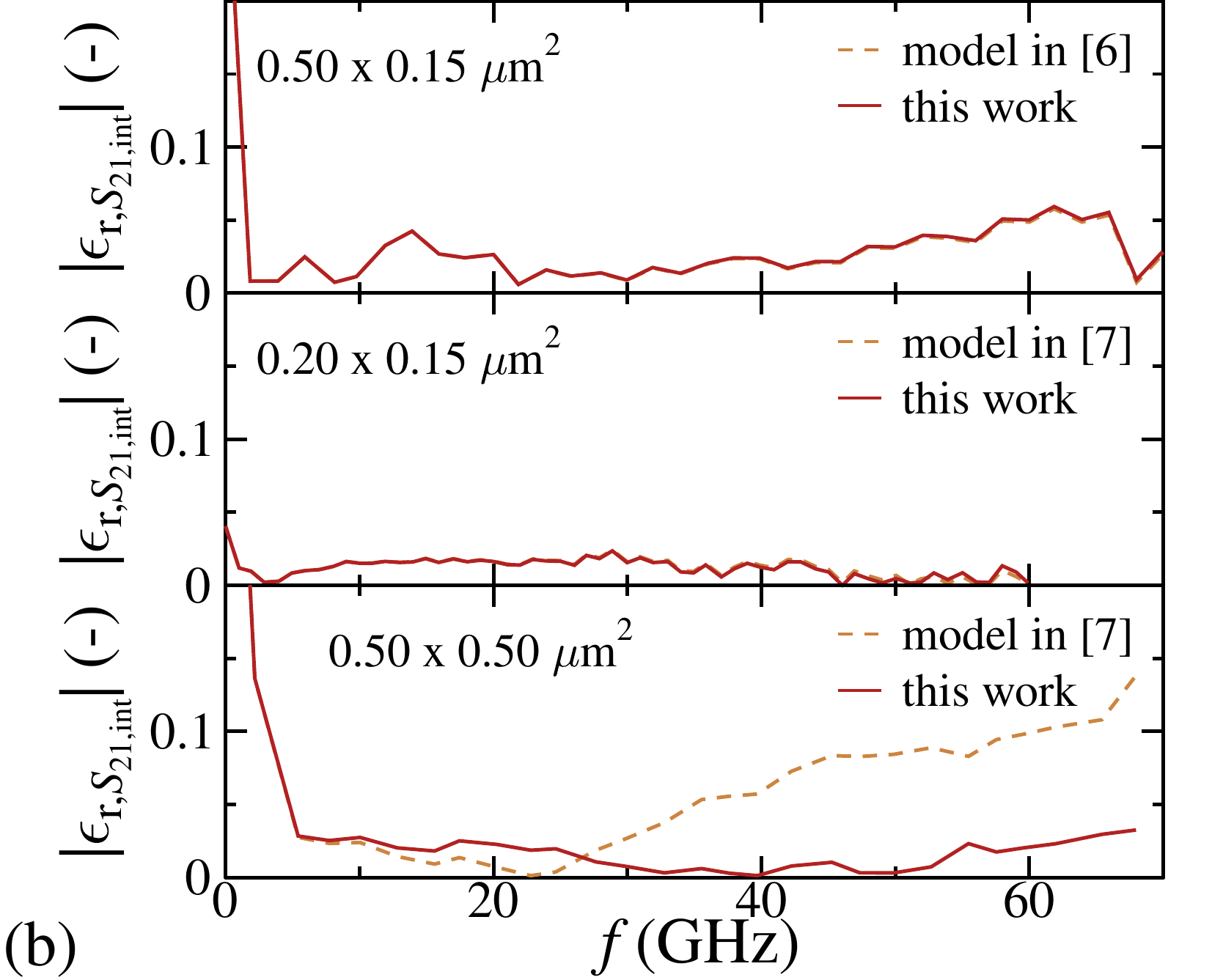}
\caption{Relative error between results of modeling approaches and experimental data of the $S_{\rm 21,int}$ at (a) ON and (b) OFF states.}
\label{fig:error}
\end{figure}

The relative error $\epsilon_{\rm r}$ of the two dynamic modeling approaches, the one with $C_{\rm ON}$ and $R_{\rm r,ON/OFF}$ (proposed here) and the oftenly used one without such elements are shown in Fig. \ref{fig:error} for $S_{\rm 21,int}$ at both states (cf. Fig. \ref{fig:results}) with respect to the experimental data.

\section*{Acknowledgements}

The authors thank Deji Akinwande from University of Texas and Enrique Miranda from Universitat Autònoma de Barcelona for fruitful discussions.

\end{document}